\documentclass[floatfix,aps,showpacs,preprintnumbers]{revtex4}
\usepackage{graphicx,epsfig}
 
\begin{document}

\title{Trojan Horse as an indirect technique in nuclear astrophysics. Resonance reactions.}

\author{A. M. Mukhamedzhanov$^1$,  L. D. Blokhintsev$^2$, B. F. Irgaziev$^3$, 
A. S. Kadyrov $^4$, M. La Cognata$^5$, C. Spitaleri$^5$ and R. E. Tribble$^1$}  
\address{$^1$ Cyclotron Institute, Texas A\&M University, College Station, Texas, 77843, USA}
\address{$^2$ Institute
 of Nuclear Physics, Moscow State University, Moscow, Russia}
\address{$^3$  Faculty of Engineering Sciences, GIK Institute of Engineering Sciences and Technology, Topi-23640,
N.W.F.P., Pakistan}
\address{$^4$   ARC Centre for Antimatter-Matter Studies, Curtin University of Technology, GPO Box U1987, Perth, WA 6845, Australia} 
\address{$^5$   DMFCI, Universit\`a  di Catania, Catania, Italy and INFN - Laboratori Nazionali del Sud, Catania, Italy}   

\begin{abstract}
The Trojan Horse method is a powerful indirect technique that provides
information to determine astrophysical factors for binary rearrangement processes $x + A \to b + B$ at astrophysically relevant energies by measuring the cross section for the Trojan Horse reaction $a + A \to y+ b + B$ in quasi-free kinematics. We present the 
theory of the Trojan Horse method for resonant binary subreactions based on the half-off-energy-shell $R$ matrix approach which takes into account the off-energy-shell effects and initial and final state interactions.
\end{abstract}

\maketitle

\section{Introduction}
The presence of the Coulomb barrier for colliding charged nuclei
makes nuclear reaction cross sections at astrophysical energies so
small that their direct measurement in the laboratory is very
difficult, or even impossible. 

Consequently indirect techniques often are used
to determine these cross sections. The Trojan Horse (TH) method is a powerful indirect technique which allows one to determine the
astrophysical factor for rearrangement reactions. 
The TH method, first suggested by Baur \cite{baur86th}, 
involves obtaining the cross section of the binary $x+ A \to b+B$ process
at astrophysical energies by measuring the two-body to three-body ($2 \to
 3$) TH process, $a + A \to
y + b+B$, in the
quasi-free (QF) kinematics regime, where the "Trojan Horse" particle,
 $a=(x\,y)$, is accelerated
at energies above the Coulomb barrier. After penetrating through the Coulomb
barrier, nucleus $a$ undergoes breakup leaving particle $x$ to interact with
 target $A$ while
projectile $y$ flies away. From the measured $a + A \to y + b+B$  cross
 section, the energy
dependence of the binary subprocess, $x+ A \to b+B$, is determined. 

The main advantage of the
TH method is that the extracted cross section of the binary subprocess does
 not contain the
Coulomb barrier factor. Consequently the TH cross section can be used to
 determine
the energy dependence of the astrophysical factor, $S(E)$, of the binary
 process, $x+ A \to b+B$,
down to zero relative
kinetic energy of the 
particles $x$ and $A$ without distortion due to electron screening
 \cite{ass87,spit01}. 
The absolute value of $S(E)$ must be found by normalization to direct
 measurements at higher
energies. At low energies where electron screening becomes
important, comparison of the astrophysical factor determined from the TH
 method to the direct
result provides a determination of the screening potential. 

Even though the TH method has been applied successfully to many direct and
 resonant processes
(see \cite{spit04} and references therein), there are still reservations
 about the reliability of the
method due to two potential modifications of the yield from off-shell
 effects and initial and final
state interactions in the TH $2 \to 3$ reaction.  Here we will address the theory of the TH 
method for resonant binary reactions $x + A \to b + B$.

\section{\bf Trojan Horse}
The TH reaction is a many-body process (at least
four-body) and its strict analysis requires many-body techniques. However some important
features of the TH method can be addressed in a simple model. Let us consider the TH process 
assuming that nuclei $y,\,x $ and $B$ are constituent particles, i. e. we neglect their internal 
degrees of freedom. For simplicity, we disregard the spins of the particles. 
The TH reaction amplitude is given in the post form by 
\begin{equation}
{\tilde M}(P, {\rm {\bf k}}_{aA}) =  <\chi_{ {\rm {\bf k}}_{yF} }^{(-)}\,\Phi_{F}^{(-)}|\Delta V_{yF}|\Psi_{i}^{(+)}>.
\label{threactampl1}
\end{equation}
Here,  $\Psi_{i}^{(+)}$ is the exact $a + A$ scattering wave function, $\Phi_{F}^{(-)}$ is the wave function of the system $F= b + B= x+A$, $\,\chi_{ {\rm {\bf k}}_{yF} }^{(-)}({\rm {\bf r}}_{ij})$ is the distorted wave of the system $y + F$, $\,\varphi_{i}$ is the bound state wave function
of nucleus $i$,  ${\rm {\bf r}}_{ij}$ and ${\rm {\bf k}}_{ij}$ are the relative coordinate and 
relative momentum of nuclei $i$ and $j$, $\,P=\lbrace {\rm {\bf k}}_{yF}, {\rm {\bf k}}_{bB} \rbrace$ is the six-dimesional momentum describing the three-body system $y,\,b$ and $B$ in the final system, $\,\Delta V_{yF}= V_{yF} - U_{yF}$, 
$\,V_{yF}= V_{yb} +V_{yB}=V_{yx} + V_{yA}$ is the interaction potential of $y$ and the system 
$F$ and $U_{yF}$ is their optical potential. The surface approximation suggested in \cite{typel03} was the first serious attempt to address the theory of the TH method.  The surface approximation assumes that the TH reaction amplitude has contributions from the external region where the interaction between the fragments $b$ and $B$ ($x$ and $A$)
can be neglected and the wave function $\Phi_{F}^{(-)}$ can be replaced 
by its leading asymptotic form 
\begin{equation}
\Phi_{F}^{(+)} \approx \varphi_{b}\,[e^{ i\,{\rm {\bf k}}_{bB}\cdot {\rm {\bf r}}_{bB}  } 
+ F_{bB}\,u^{(+)}_{k_{bB}}(r_{bB})]  + \sqrt{\frac{v_{bB}}{v_{xA}} } M_{bB \to xA}\,\frac{ 1 }{2\,i\,k_{bB}}\,u_{k_{xA}}^{(+)}(r_{xA}),
\label{asymptphif1}
\end{equation}
where $\Phi_{F}^{(+)} \equiv \Phi^{(+)}_{ {\rm {\bf k}}_{bB}(F)}$ and  $\Phi^{(-)}_{{\rm {\bf k}}_{bB}(F)}= \Phi^{(+)*}_{-{\rm {\bf k}}_{bB}(F)}$, $u_{k_{ij}}^{(+)}(r_{ij})$ is the outgoing spherical wave, 
$F_{bB}$ is the $b + B$ elastic scattering 
amplitude,  $M_{bB \to xA}$ is the $b + B \to x + A$ reaction amplitude inverse to the binary reaction $ x + A \to b + B$ and $v_{ij}$ is the relative velocity of nuclei $i$ and $j$. 
The expression for the TH reaction amplitude in the surface approximation is given by
\begin{equation}
{\tilde M}(P, {\rm {\bf k}}_{aA}) \sim  M_{bB \to xA}\, <\chi_{ {\rm {\bf k}}_{yF} }^{(-)}\,\varphi_{A}\,u_{k_{xA}}^{(-)}(r_{xA})|\Delta V_{yF}|\varphi_{a}\,\varphi_{A}\,\chi_{{\rm {\bf k}}_{aA}}^{(+)}({\rm {\bf r}}_{aA} ) >,
\label{threactampl2}
\end{equation}
where the exact initial scattering wave function $\Psi_{i}^{(+)}$ is replaced by $\varphi_{a}\,\varphi_{A}\,\chi_{{\rm {\bf k}}_{aA}}^{(+)}({\rm {\bf r}}_{aA} )$ and $\chi_{aA}^{(+)}$ is the distorted wave describing the scattering of the nuclei $a$ and $A$ in the initial state of the TH reaction. For simplicity we don't take into account here the Coulomb interactions. 
However, in the case of the resonant binary reaction $x + A \to b + B$ the dominant contribution comes from the nuclear interior where both channels $x + A$ and $ b + B$ are coupled and where the asymptotic approximation for $\Phi_{F}^{(+)}$ cannot be applied\footnote{Generally speaking one must be very careful in using the asymptotic approximation for the scattering wave function  $\Phi_{F}^{(-)}$ because the matrix element with the exact wave function in the initial state and ingoing spherical wave $u_{k_{xA}}^{(-)}(r_{xA})$  in the final state vanishes after transformation of the volume integral into a surface integral \cite{kadyrov}.}. 

In this work we will address the theory of the TH method for the resonant binary subprocesses $x + A \to b +B$ which explicitly takes into account the off-shell character of $x$. 
Eq. (\ref{threactampl1}) can be used as a starting point to derive the expression for the TH reaction amplitude.
We assume that the resonant reaction $x + A \to b + B$  proceeds through the formation of the intermediate compound state $ \Phi_{i}$, i. e. we neglect the direct coupling between the initial $x+ A$ 
and final $b + B$ channels, which contributes dominantly to direct reactions but gives negligible contribution to resonant ones. An important step in deriving the resonant contribution to the TH reaction matrix element is the spectral decomposition for the wave function $\Phi_{F}^{(-)}$ given by Eq. (3.8.1) \cite{mahaux}. It leads to the shell-model based resonant R matrix representation for $\Phi_{F}^{(-)}$ which is similar to the level decomposition for the wave function in the internal region 
in the $R$ matrix approach:
\begin{equation}
\Phi _F^{( - )} \approx \sum\limits_{\nu,\tau = 1}^N {{\tilde V}_\nu^{bB}(E_{bB})\,[{\bf D}^{ - 1} } ]_{\nu \tau}\,\Phi_{\tau}.
\label{spectrdecpsi2}
\end{equation}
Here $N$ is the number of the levels included, $E_{bB}$ is the relative kinetic energy of nuclei $b$ and $B$, 
$\Phi _\tau$ is the bound state wave function describing the compound system $F$ excited to the level $\tau$. 
$D_{\nu \tau}$ is similar to the level matrix in the $R$ matrix theory and is given by Eq. (4.2.20b) \cite{mahaux}.
Finally, 
\begin{equation}
{\tilde V}_\nu^{bB} (E_{bB})= <\chi_{bB}^{(-)}\,\varphi_{b}|\Delta V_{bB}|\Phi_{\nu}>
\label{resformfactr1}
\end{equation}
is the resonant form factor for the decay of the resonance $F_{\nu}$ described by the compound state $\Phi_{\nu}$ into the channel $b + B$. The partial resonance width is given by 
\begin{equation}
{\tilde \Gamma}_{\nu}(E_{bB})= 2\,\pi|{\tilde V}_\nu^{bB}(E_{bB} )|^{2}. \label{reswidth1}
\end{equation}
Then the TH reaction amplitude is
\begin{eqnarray}
{\tilde M}^{(R)}(P, {\rm {\bf k}}_{aA}) \approx \sum\limits_{\nu,\tau = 1}^N\,{{\tilde V}_{\nu}^{bB} (E_{bB})\,[{\bf D}^{ - 1}]_{\nu \tau} }\,{\tilde M}_{\tau}({\rm {\bf k}}_{yF}, {\rm {\bf k}}_{aA}), 
\label{exthreactampl1} 
\end{eqnarray}  
where ${\tilde M}_{\tau}({\rm {\bf k}}_{yF}, {\rm {\bf k}}_{aA})$ is the exact amplitude for the direct transfer reaction $a + A \to y + F_{\tau}$ populating the compound state $F_{\tau}$ of the system $F=x + A = b + B$: 
\begin{equation}
{\tilde M}_{\tau}({\rm {\bf k}}_{yF}, {\rm {\bf k}}_{aA})= < \chi_{yF}^{(-)}\,\Phi_{\tau}|\Delta V_{yF}|\Psi_{i}^{(+)}>.
\label{extrampl1}
\end{equation}
The direct transfer reaction is very well described by the DWBA amplitude, i. e. for the practical analysis we can approximate  $\Psi_{i}^{(+)} \approx \varphi_{a}\,\varphi_{A}\,\chi_{aA}^{(+)}$.
Correspondingly, ${\tilde M}_{\tau}({\rm {\bf k}}_{yF}, {\rm {\bf k}}_{aA})$ can be replaced 
by 
\begin{equation}
{\tilde M}_{\tau}^{DW}({\rm {\bf k}}_{yF}, {\rm {\bf k}}_{aA})= < \chi_{yF}^{(-)}\,\Phi_{\tau}|\Delta V_{yF}|\varphi_{a}\,\varphi_{A}\,\chi_{i}^{(+)}>.
\label{dwtrampl1}
\end{equation}
Correspondingly for the TH reaction amplitude we get from  Eq. (\ref{exthreactampl1})
\begin{eqnarray}
{\tilde M}^{(R)}(P, {\rm {\bf k}}_{aA}) \approx \sum\limits_{\nu,\tau = 1}^N\,{{\tilde V}_\nu^{bB} (E_{bB})\,[{\bf D}^{ - 1}]_{\nu \tau} }\,{\tilde M}^{DW}_{\tau}({\rm {\bf k}}_{yF}, {\rm {\bf k}}_{aA}). 
\label{threactampldw1} 
\end{eqnarray}  
The DWBA amplitude takes into account the rescattering of nuclei $a$ and $A$ in the initial state of the TH reaction and enters as a form factor into the TH resonant reaction amplitude reflecting the off-energy shell   character of the transferred particle $x$.
Since in the TH method the astrophysical factor determined from the TH method is normalized 
to the on-energy-shell (OES) $S$ factor, the replacement of the exact transfer amplitude by the DWBA one, as we will see, practically does not affect the final result. 

\subsection{Single resonance}

The triple differential cross section for the TH process  $a + A \to y+ b +B$ 
proceeding through an isolated resonance $F_{\tau}$ is given by 
\begin{eqnarray}
\frac{ {\rm d}^3 \sigma  }{ {\rm d}E_{bB}\,{\rm d}\Omega _{ {\rm {\bf k}}_{bB} }\,{\rm d}\Omega _{{\rm {\bf k}}_{yF} } }= \lambda_{3}\, \frac{\Gamma_{bB(\tau)}(E_{bB})\,
|M^{DW}_{\tau}({\rm {\bf k}}_{yF},{\rm {\bf k}}_{aA})|^2 }{ (E_{xA} - E_{R_{\tau}})^{2} + \frac{\Gamma_{\tau}
^{2}(E_{xA})}{4}}. 
\label{trdifcrsectm1}
\end{eqnarray}
Here, $\lambda_{3}$ is the kinematical factor, $\Gamma_{bB(\tau)} (E_{bB})$ is the observable resonance partial width in the channel $b+B$, $\,\Gamma_{\tau}(E_{xA})$ is the total observable width of the resonance $F_{\tau}$.
Note that all functions $T(E)$ are related to ${\tilde T}(E)$ as $T(E)={\tilde T}(E)/(1- (\frac{{\rm d}\,\Delta_{\tau \tau}}{{\rm d}E})_{E=E_{R_{\tau}}})$, where $\Delta_{\tau \tau}$ is the $\tau$ level shift.
Also  $E_{R_{\tau}}$ is the resonance energy of the resonance $F_{\tau}$ in the channel $x + A$. 
Thus the TH triple differential cross section, in contrast to the OES single-level resonance cross section,  contains the generalized form factor  $|M_{\tau}^{DW}({\rm {\bf k}}_{yF},{\rm {\bf k}}_{aA})|^{2}$ rather 
then the entry channel partial resonance width $\Gamma_{xA(\tau)}(E_{xA})$  of the
binary process $x + A \to b + B$. A simple renormalization of the TH triple differential cross section allows us to single out the OES astrophysical factor for the resonant binary subprocess $x + A \to b + B$:
\begin{eqnarray}
S(E_{xA}) = NF(E_{xA})\, \frac{ {\rm d}^3 \sigma  }{ {\rm d}E_{bB}\,{\rm d}\Omega _{ {\rm {\bf k}}_{bB} }\,{\rm d}\Omega _{{\rm {\bf k}}_{yF} } }  
= \frac{\pi}{2\,\mu_{xA}}\,e^{2\,\pi\,\eta_{xA}}\,\frac {\Gamma_{bB(\tau)}(E_{bB})\,
\Gamma_{xA(\tau)}(E_{xA})}{(E_{xA}  - E_{R_{\tau}})^{2} +  
\frac{\Gamma_{\tau}^{2}(E_{xA})}{4}},
\label{doubldiffcrsect1} 
\end{eqnarray}
where the normalization factor $NF(E_{xA})$ is given by
\begin{equation}
NF(E_{xA})=\frac{\pi}{k_{xA}^{2}}\,\frac{1}{\lambda_{3}}\,E_{xA}\,
e^{2\,\pi\,\eta_{xA}}\,
\frac{\Gamma_{xA(\tau)}(E_{xA})}{ |M_{\tau}^{DW}({\rm {\bf k}}_{yF},{\rm {\bf k}}_{aA})|^2}.
\label{normfctr1}
\end{equation}
Note that the DWBA amplitude $M^{DW}({\rm {\bf k}}_{yF},{\rm {\bf k}}_{aA})$ remains practically constant on the interval of a few hundreds keV. Eq. (\ref{doubldiffcrsect1}) explaines and justifies the phenomenological procedure used before successfully in the TH analysis
(see \cite{spit04} and references therein).  The renormalization factor can be rewritten as 
\begin{equation}
NF(E_{xA})= e^{2\,\pi\,\big[\eta_{xA} - \eta_{xA}\big|_{ E_{xA}=E_{R_{1}} } \big]}\,
\frac{\Gamma_{xA(\tau)}(E_{xA})}{\Gamma_{xA(\tau)}(E_{R_{1}}) }\,NF(E_{R_{1}}),
\label{normfctr2}
\end{equation}
where $\Gamma_{xA(\tau)}(E_{xA})/\Gamma_{xA(\tau)}(E_{R_{1}}) = P(E_{xA})$ is the barrier penetration factor appearing in the $R$ matrix theory. 
The factor $NF(E_{R_{1}})$ can be found phenomenologically by comparing the experimental TH triple differential cross section with the available OES experimental astrophysical factor at resonance energy. This phenomenological 
normalization leads to the intermediate  astrophysical factor
\begin{eqnarray}
S'(E_{xA}) =   \frac{\pi}{2\,\mu_{xA}}\,e^{2\,\pi\,\eta_{xA}|_{E_{xA}=E_{R_{1}}} }\,\frac {\Gamma_{bB(\tau)}(E_{bB})\,\Gamma_{xA(\tau)}(E_{R_{1}})}{(E_{xA}  - E_{R_{\tau}})^{2} +  \frac{\Gamma_{\tau}^{2}(E_{xA})}{4}}.
\label{intermsfctr1} 
\end{eqnarray}
The final astrophysical factor can be derived by multiplying $S'(E_{xA})$ by the energy-dependent factor in 
Eq. (\ref{normfctr2})  $e^{2\,\pi\,\big[\eta_{xA} - \eta_{xA}\big|_{ E_{xA}=E_{R_{1}} }\big ]}\,\Gamma_{xA(\tau)}(E_{xA})/\Gamma_{xA(\tau)}(E_{R_{1}})$. 
Thus normalization of the triple TH differential cross section to the experimental astrophysical factor at resonance energy achieved by multiplying Eq. (\ref{trdifcrsectm1}) by the factor $NF(E_{xA})$ plays a very special role in the TH method. 

\subsection{Two interfering resonances}

For two interfering resonances we need to consider the two-level, two channel case. This requires the half-off-energy-shell (HOES) $R$ matrix formalism. Here we address this formalism for a simple case when the distances between two resonances are significantly larger then their total widths. Then the OES reaction amplitude
in the $R$ matrix formalism is given by the sum of the amplitude of each resonances (see Eq. (XII,5.15) \cite{thomaslane58}). The corresponding expression for the HOES reaction amplitude can be obtained by the replacement of the resonance partial widths in the entry channel of the binary reaction $x + A \to b + B$ by 
the corresponding generalized form factors $M_{\tau}^{DW}({\rm {\bf k}}_{yF},{\rm {\bf k}}_{a}),\,\,\tau=1,2$.
Thus the triple TH cross section in the presence of two interfering resonances in the subsystem 
$F=x+ A=b + B$ is given by
\begin{eqnarray}
\frac{{ {\rm d}^2 \sigma }}{{ {\rm d}E_{bB}\,{\rm d}\Omega _{ {\rm {\bf {k}}}_{yF} }}\,{\rm d}\Omega _{ {\rm {\bf {k}}}_{bB} }} 
=\lambda_{3}\,\big|\sum\limits_{\tau=1,2} \, \frac {\Gamma_{bB(\tau)}^{1/2}(E_{bB})\,M_{\tau}^{DW}({\rm {\bf k}}_{yF},{\rm {\bf k}}_{aA})}{E_{xA}  - E_{R_{\tau}} + i \,\frac{\Gamma_{\tau}(E_{xA})}{2}}\big|^{2}.  
\label{tworesonances1}
\end{eqnarray}
We assume that $E_{R_{1}} < E_{R_{2}}$. The goal of the THM
is to determine the energy dependence of the astrophysical factor at the astrophysically relevant energies.
The ratio 
$M_{21}^{DW}= M_{2}^{DW}({\rm {\bf k}}_{yF},{\rm {\bf k}}_{aA})/M_{1}^{DW}({\rm {\bf k}}_{yF},{\rm {\bf k}}_{aA})$ is practically constant in the interval of a few hundred keV, $E_{xA} \leq  E_{R_{1}}$. 
Normalizing the TH cross section to the OES $S$ factor at $E=E_{R_{1}}$, where the contribution from the second resonance can be neglected, 
gives the astrophysical factor determined from the TH reaction
\begin{eqnarray}
 S^{TH}(E_{xA})= \frac{\pi\,e^{2\,\pi\,\eta_{xA}}}{2\mu_{xA}}\,\Gamma_{xA(1)}(E_{xA})\,
\big|\big \lbrack \frac {\Gamma_{bB(1)}^{1/2}(E_{bB})}{E_{xA}  - E_{R_{1}} + i \,\frac{\Gamma_{1}(E_{xA})}{2}}  +  \frac {\Gamma_{bB(2)}^{1/2}(E_{bB})\,M_{21}^{DW}}{E_{xA}  - E_{R_{2}} + i \,\frac{\Gamma_{\tau}(E_{xA})}{2}} \big \rbrack \big|^{2}. 
\label{Sfctrtworesonances1}
\end{eqnarray}
This astrophysical factor is to be compared with the OES astrophysical factor determined from direct measurements
\begin{eqnarray}
S(E_{xA}) = \frac{\pi\,e^{2\,\pi\,\eta_{xA}}}{2\mu_{xA}}\,\Gamma_{xA(1)}(E_{xA})\,\big|\big \lbrack \frac {\Gamma_{bB(1)}^{1/2}(E_{bB})}{E_{xA}  - E_{R_{1}} + i \,\frac{\Gamma_{1}(E_{xA})}{2}}  
+  \frac {\Gamma_{bB(2)}^{1/2}(E_{bB})\,\gamma_{(xA)21}}{E_{xA}  - E_{R_{2}} + i \,\frac{\Gamma_{\tau}(E_{xA})}{2}} \big \rbrack \big|^{2}. 
\label{SfctrtwoR2}
\end{eqnarray}
Here, $\gamma_{(xA)21}=\gamma_{(xA)2}/\gamma_{(xA)1}= \Gamma_{xA(2)}^{1/2}(E_{xA})/\Gamma_{xA(1)}^{1/2}(E_{xA})$ and $\,\gamma_{(xA){\tau}}$ is the reduced width for the $\tau$-th resonance in the channel $x + A$. Each amplitude $M_{2}^{DW}({\rm {\bf k}}_{yF},{\rm {\bf k}}_{a})$ is complex, but the ratio $M_{21}^{DW}$ may have a small imaginary part. The normalization of the TH $S$ factor to the OES one at resonance energy plays a crucial role in the TH method. After such a normalization, we need to know only the ratio of the DWBA
amplitudes to calculate $S^{TH}(E_{xA})$. 

\subsubsection{Plane wave approximation}

\begin{figure}[!t]
\begin{center}
\includegraphics{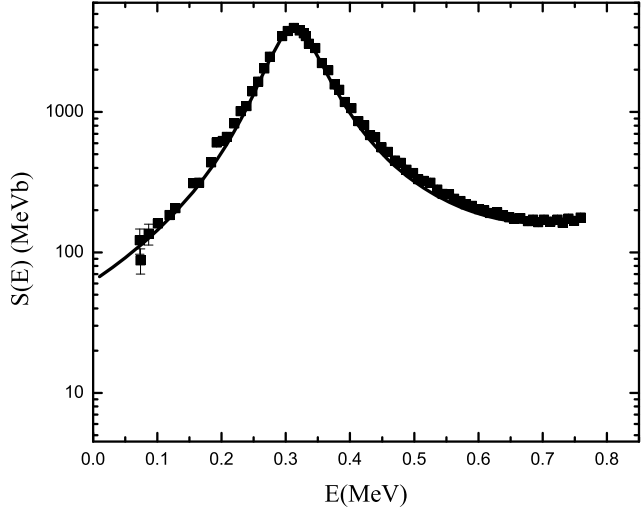}
\vspace{-0.1cm}
\caption{Comparison of the calculated astrophysical factor $S^{TH}(E)$ for ${}^{15}{\rm N}(p, \alpha){}^{12}{\rm C}$ (solid line), where $E \equiv E_{xA}$, with the direct data \cite{schardt52,zyskind79,redder82}.}
\label{fig_res}
\vspace{-0.5 cm}
\end{center}
\end{figure}

Ratio $M_{21}^{DW}$ can be approximated by the ratio of the corresponding amplitudes
calculated in a plane wave approximation, because a simple plane wave approximation gives similar angular and energy dependence as the DWBA but fails to reproduce the absolute value.
It explains why a simple plane wave approximation works well in the TH analysis \cite{spit04}.   
Note that in the plane wave approximation $M_{\tau}^{DW}({\rm {\bf k}}_{yF}, {\rm {\bf k}}_{aA})$ is replaced by
\begin{eqnarray}
M_{\tau}^{0}({\rm {\bf k}}_{yF}, {\rm {\bf k}}_{aA})= < e^{i\, {\rm {\bf k}}_{yF} \cdot  {\rm {\bf r}}_{yF}}   \varphi_{y}\,\Phi_{\tau}| V_{yA} + V_{xA}|\,\varphi_{a}\,\varphi_{A}\,e^{i\,{\rm {\bf k}}_{aA} \cdot {\rm {\bf r}}_{aA} }>.   \label{mplwavepr1}  
\end{eqnarray}
Note that the post and prior forms are equivalent but the post form is more convenient for our purpose. In the QF kinematics for sufficiently high momentum of the the projectile $A$ it will interact dominantly with the fragment $x$ while the contribution of the term with $V_{yA}$ is minimized. That is why in what follows we neglect the term containing $V_{yA}$. Then the transfer reaction amplitude in the plane wave approximation takes the form
\begin{eqnarray}
M_{\tau}^{0}({\rm {\bf k}}_{yF}, {\rm {\bf k}}_{aA}) \approx < e^{i\, {\rm {\bf k}}_{yF} \cdot  {\rm {\bf r}}_{yF}}\,I^{F_{\tau}}_{xA}| <V_{xA}>_{xA}|\,I^{a}_{yx}\,e^{i\,{\rm {\bf k}}_{aA} \cdot {\rm {\bf r}}_{aA} }>, 
\label{mplwavepr2}
\end{eqnarray}
where 
$\,I^{F_{\tau}}_{xA}=<\varphi_{A}\,\varphi_{x}|\Phi_{\tau}>$ is the
overlap function of the wave function of the resonance state $F_{\tau}$ and the bound state wave functions of $A$ and $x$,  $\,I^{a}_{yx}=<\varphi_{y}\,\varphi_{x}|\varphi_{a}>$  is the overlap function of the bound state wave functions of nuclei $a,\,x$ and $y$, and $\varphi_{x}$,  $\,\,<V_{xA}> = <\varphi_{A}\,\varphi_{x}|V_{xA}|\varphi_{x}\,\varphi_{A}>$.  The plane wave amplitude $M_{\tau}^{0}({\rm {\bf k}}_{yF}, {\rm {\bf k}}_{aA})$
can be written in a factorized form
\begin{equation}
M_{\tau}^{0}({\rm {\bf k}}_{yF}, {\rm {\bf k}}_{aA}) =  [W^{F_{\tau}}_{xA}({\rm {\bf k}}_{A} - \frac{m_{A}}{m_{F}}\,{\rm {\bf k}}_{F})]^{*}\,I^{a}_{yx}({\rm {\bf k}}_{y} - \frac{m_{y}}{m_{a}}\,{\rm {\bf k}}_{a}). 
\label{mplwavefact1}
\end{equation}
Here, $I^{a}_{yx}({\rm {\bf p}}_{yx})$ is the Fourier transform of the 
overlap function $I^{a}_{yx}({\rm {\bf r}}_{yx})$ and 
\begin{equation}
W^{F_{\tau}}_{xA}({\rm {\bf k}}_{xA})
=<e^{i\, {\rm {\bf k}}_{xA} \cdot  {\rm {\bf r}}_{xA}}|<V_{xA}>_{xA}({\rm {\bf r}}_{xA})|I^{F_{\tau}}_{xA}
({\rm {\bf r}}_{xA})>
\label{vertfrmfactr1}
\end{equation}
is the vertex form factor for $x + A \to F_{\tau}$.
Then Eq. (\ref{tworesonances1}) for the TH triple differential cross section takes the form
\begin{widetext}
\begin{eqnarray}
\frac{{ {\rm d}^2 \sigma }}{{ {\rm d}E_{bB}\,{\rm d}\Omega _{ {\rm {\bf {k}}}_{yF} }}\,{\rm d}\Omega _{ {\rm {\bf {k}}}_{bB} }} 
 = \lambda_{3}\,|I^{a}_{yx}({\rm {\bf k}}_{y} - \frac{m_{y}}{m_{a}}\,{\rm {\bf k}}_{a})|^{2}\, \big|\sum\limits_{\tau=1,2} \, \frac {\Gamma_{bB(\tau)}^{1/2}(E_{bB})\,[W^{F_{\tau}}_{xA}({\rm {\bf k}}_{A} - \frac{m_{A}}{m_{F}}\,{\rm {\bf k}}_{F})]^{*}}{E_{xA}  - E_{R_{\tau}} + i \,\frac{\Gamma_{\tau}(E_{xA})}{2}}\big|^{2}.   \label{tworesonances11}
\end{eqnarray}
\end{widetext}
Now we can get the HOES cross section for the binary subprocess $x + A \to b + B$  from the triple differential cross section 
\begin{eqnarray}
\left(\frac{d\sigma}{ d\Omega_{c.m.}}\right)^{\rm HOES}
\propto
\left[\frac{{ {\rm d}^2 \sigma }}{{ {\rm d}E_{bB}\,{\rm d}\Omega _{ {\rm {\bf {k}}}_{yF} }}\,{\rm d}\Omega _{ {\rm {\bf {k}}}_{bB} }} \right]\,\frac{1}{\lambda_{3} \;
|I^{a}_{yx}({\rm {\bf p}}_{yx})|^2 },  
\label{tbthreebodycrsect1}
\end{eqnarray}
where ${\rm {\bf p}}_{yx}={\rm {\bf k}}_{y} - \frac{m_{y}}{m_{a}}\,{\rm {\bf k}}_{a}$.
Eq. (\ref{tbthreebodycrsect1}) explains and justifies the procedure used in IA \cite{spit04} to connect the triple and binary TH cross sections. Note that in a strict approach the triple differential cross section is expressed in terms of the overlap function $I^{a}_{yx}$ rather then the two-body bound state wave function $\varphi_{a}$. Note that $I^{a}_{yx}$ and $\varphi_{a}$ are related by 
\begin{equation}
I^{a}_{yx}= S^{1/2}_{yx}\,\varphi_{a}, 
\label{ovrlpfunctwf1}
\end{equation}
where $S^{1/2}_{yx}$ is the spectroscopic factor.
The binary reaction HOES cross section is only intermediate result. The final goal is the TH astrophysical 
factor which can be determined by normalization of the triple differential cross section 
to the OES astrophysical factor in the first resonance peak  and is given by Eq. (\ref{Sfctrtworesonances1}).
In the plane wave approximation $M_{21}^{DW}$ is replaced by
\begin{equation}
M_{21}^{0} = \frac{[W^{F_{\tau}}_{xA}({\rm {\bf k}}_{A} - \frac{m_{A}}{m_{F}}\,{\rm {\bf k}}_{F})]^{*}}
{[W^{F_{\tau}}_{xA}({\rm {\bf k}}_{A} - \frac{m_{A}}{m_{F}}\,{\rm {\bf k}}_{F})]^{*}}.
\label{m21plwave1} 
\end{equation}
If $M_{21}^{0} \approx \gamma_{(xA)21}$, the astrophysical factor $S^{TH}(E_{xA})$ reproduces the OES $S$ factor $S(E_{xA})$ at energies $E_{xA} \leq E_{R_{1}}$.   
In Fig. \ref{fig_res} the astrophysical factor $S^{TH}(E_{xA})$ for ${}^{15}{\rm N}(p, \alpha){}^{12}{\rm C}$ calculated using Eq. (\ref{Sfctrtworesonances1}) for the TH reaction ${}^{15}{\rm N}(d, n\,\alpha){}^{12}{\rm C}$ is compared with the experimental $S(E_{xA})$ obtained from direct measurements. There are two $1^{-}$ interfering resonances at $E_{R_{1}}=312$ keV and $E_{R_{2}}=962$ keV. The best fit has been achieved for  $\Gamma _{xA(1)} \equiv \Gamma_{p(1)} = 1.1 $ keV, $\,\Gamma _{bB(1)} \equiv \Gamma _{\alpha(1)}= 93.4$ keV,  $\Gamma_{xA(2)} \equiv \Gamma_{p(2)} = 95.31$ keV and $\Gamma _{bB(2)} \equiv \Gamma _{\alpha(2)} = 45$ keV. 
 To find $M_{21}^{0}$ we used Eq. (\ref{vertfrmfactr1}) in which the overlap function $I^{F_{(i)}}_{xA}$ is approximated by a single-particle ${}^{15}{\rm N}-p$ wave function in the Woods-Saxon potential  
calculated in the internal region by a procedure similar to that used in R-matrix
method to calculate the level eigenfunctions. We find that $M_{21}^{0} \approx 1.13$ while $\gamma_{(xA)21}= 1.1 \pm 0.1$. It explains why the calculated $S^{TH}(E_{xA})$ shown in Fig. \ref{fig_res} is  in an excellent agreement with the direct data. \\
We presented the expression for the resonant $S$ factor determined from the TH reaction taking into account the off-energy-shell effects  within the HOES $R$ matrix formalism and justified a simple plane wave approximation. Validating this makes it clear why the TH method is such a
powerful indirect technique for nuclear astrophysics. 

This work was supported in part by the U.\,S. DOE under Grant No.\@
 DE-FG02-93ER40773.

\end{document}